\documentclass[aps, pra, reprint, longbibliography]{revtex4-2}
\usepackage{graphicx} 
\usepackage{amsmath, physics}
\usepackage{amssymb}
\usepackage{amsfonts}
\usepackage[unicode]{hyperref}
\hypersetup{
   unicode=true,          %
   plainpages=false,
   colorlinks=true,       
   citecolor=blue,        
   urlcolor=blue
}
\usepackage[caption=false,position=top,singlelinecheck=off,justification=raggedright]{subfig}
\usepackage{color}

\begin{document}
\title{Universality of dissipative discrete time crystal formation}

\author{Roy D. Jara Jr.}
\email{rjara@nip.upd.edu.ph}
\affiliation{National Institute of Physics, University of the Philippines, Diliman, Quezon City 1101, Philippines}

\author{Jayson G. Cosme}
\email{jcosme@nip.upd.edu.ph}
\affiliation{National Institute of Physics, University of the Philippines, Diliman, Quezon City 1101, Philippines}


\begin{abstract}

We demonstrate that the Kibble-Zurek mechanism (KZM) holds for open systems transitioning from a disordered phase to a discrete time crystal (DTC). Specifically, we observe the characteristic power-law scaling with quench time of the number of spatial defects and the transition delay measured from the time at which the system crosses the critical point. We show analytically that this universal behavior can be traced back to how systems that can be mapped onto a dissipative linear parametric oscillator (DLPO) satisfy the adiabatic-impulse (AI) approximation, evinced by the divergence of the relaxation time of the DLPO near a critical point. We verify our predictions in both the classical and quantum regimes by considering two systems: the Sine-Gordon model, which is a paradigmatic system for emulating classical DTCs; and the open Dicke lattice model, an array of spin-boson systems subject to quantum fluctuations. We establish a universality class for DTC formation in systems that can be mapped onto a DLPO and show that the classical and quantum models considered here belong to this class.

\end{abstract}

\maketitle

\section{Introduction}

The KZM, originally formulated in a cosmological context \cite{kibble_implications_1980, zurek_cosmological_1985, zurek_cosmological_1996, ruutu_vortex_1996}, is a widely celebrated theory in understanding the universal behavior of quenched systems. It relates the defect formation on systems transitioning from a disordered phase with the speed at which a critical point is crossed, and the critical exponents associated with the systems' universality class \cite{del_campo_universality_2014}. The KZM has been tested across multiple platforms, ranging from closed systems \cite{Lamporesi_KZM_BEC_2013, Chomaz_KZM_2DBEC_2015, griffin_scaling_2012, ye_universal_2018, liu_kibble-zurek_2020, navon_critical_2015, nagy_self-organization_2008, shimizu_dynamics_2018, dziarmaga_tensor_2023, anquez_quantum_2016, schmitt_quantum_2022, li_probing_2023, du_kibblezurek_2023, keesling_quantum_2019, chepiga_kibble-zurek_2021, cui_experimental_2016, ulm_observation_2013, pyka_topological_2013, yuan_kibble-zurek_2024}, to driven \cite{reichhardt_kibble-zurek_2022, maegochi_kibble-zurek_2022, yang_universal_2023, clark_universal_2016, russomanno_kibble-zurek_2016, anderson_direct_2017, zamani_scaling_2024, verstraelen_classical_2020, sadeghizade_anti_kzm_2025}, and open systems \cite{laguna_density_1997, laguna_critical_1998, suzuki_deconstructing_2025, rossini_dynamic_2020, zamora_kibble-zurek_2020, puebla_universal_2020, bacsi_kibblezurek_2023, klinder_dynamical_2015, zeng_universal_2023, jara_jr_apparent_2024, kou_kibble-zurek_2025}.

Despite its success in explaining the proliferation of defects in quenched systems, the KZM has so far only been applied in systems undergoing a transition towards a spatially ordered phase that remains static in the long-time limit. Thus, the question remains whether the KZM also applies to systems undergoing a spatiotemporal ordering, for instance, in DTCs. A DTC is a dynamical phase of many-body systems that emerges from a spontaneously broken discrete time translation symmetry imposed by an external periodic drive \cite{else_discrete_2020, zaletel_colloquium_2023}. It is characterized by subharmonic oscillation of observables with a period slower than the drive, wherein the most fundamental example is period-doubled DTCs \cite{else_discrete_2020, zaletel_colloquium_2023}. The DTCs have been extensively studied and observed experimentally across multiple platforms, ranging from, but not limited to, networks of classical oscillators \cite{heugel_classical_2019, yao_classical_2020, nicolaou_anharmonic_2021, heugel_role_2023, yi-thomas_theory_2024}, spin systems \cite{FloquetLMG, euler_metronome_2024, Lazarides_2020, frey_realization_2022, Floquet, pizzi_higher-order_2021, chinzei_criticality_2022, zhang_subexponential_2023, osullivan_signatures_2020, liu_discrete_2023}, bosonic systems \cite{Pizzi, bakker_driven-dissipative_2022, kongkhambut_2021, skulte_parametrically_2021, kesler_observation_2021, tuquero_dissipative_2022, cosme_time_2019, taheri_all-optical_2022}, superconductors \cite{Ojeda_2023, ojeda_emergent_2021, homann_higgs_2020}, and particles under oscillating mediums \cite{kuros_phase_2020, giergiel_creating_2020, simula_droplet_2023}.

While it has been shown that DTCs can exhibit critical behaviors reminiscent of those observed in equilibrium systems, such as roton mode softening \cite{nie_mode_2023}, and the divergence of the relaxation time \cite{yi-thomas_theory_2024, chinzei_criticality_2022, zhang_subexponential_2023} and fluctuations \cite{yi-thomas_theory_2024}, it remains to be seen whether a quench leading to the formation of a DTC exhibits universal behaviors. In this work, we demonstrate that the KZM holds for open systems transitioning from a disordered to a spatiotemporally ordered phase. To this end, we show that the AI approximation, which is the cornerstone of the KZM, applies to a DLPO. A key implication of our work is that if a spatially extended system can be mapped onto a DLPO, then it should exhibit the main signatures of the KZM following a quench that leads to a DTC formation. These signatures are the power law scaling with the quench time $\tau_{q}$ of the defect number and the transition delay from the time the system crosses the critical point. We first test this prediction on the Sine-Gordon model (SGM), which is a paradigmatic model for studying classical DTCs \cite{yao_classical_2020}. We then consider the open Dicke lattice model (DLM), which describes a lattice of spin-boson systems subject to quantum fluctuations due to cavity dissipation \cite{zou_implementation_2014}. By considering these two systems, we verify the universality of the KZM on dynamical ordered phases both in the classical and quantum regimes. We propose a universality class for DTC-forming systems and show that both the SGM and the DLM fall under this class. Thus, we generalize the notion of universality to include dynamical phases.

The paper is structured as follows. In Sec.~\ref{sec:ai_approx_dlpo}, we revisit the theory for the AI approximation, and demonstrate how this can be used to infer the dynamics of the DLPO being quenched towards a period-doubling dynamics. Using this, we formulate the main claim of this work, which is the existence of a universality for DTC-forming systems that can be probed using the KZM. In Sec.~\ref{sec:system}, we introduce the two systems that we have considered: the SGM and the DLM. We then present in Sec.~\ref{sec:kzm_dtc} our main results, which are the signatures of universality in DTC-forming systems through the power-law scaling of the transition delay, defect number, and correlation length, in accordance with the KZM. We verify in Sec.~\ref{sec:universality} that the DTCs in the SGM and the DLM belong to the same universality class associated with the DLPO. Finally, we provide a summary and a possible extension of our work in Sec.~\ref{sec:conclusion}.

\section{AI approximation for periodically driven systems}\label{sec:ai_approx_dlpo}

\begin{figure}
    \centering
    \includegraphics[scale = 0.32]{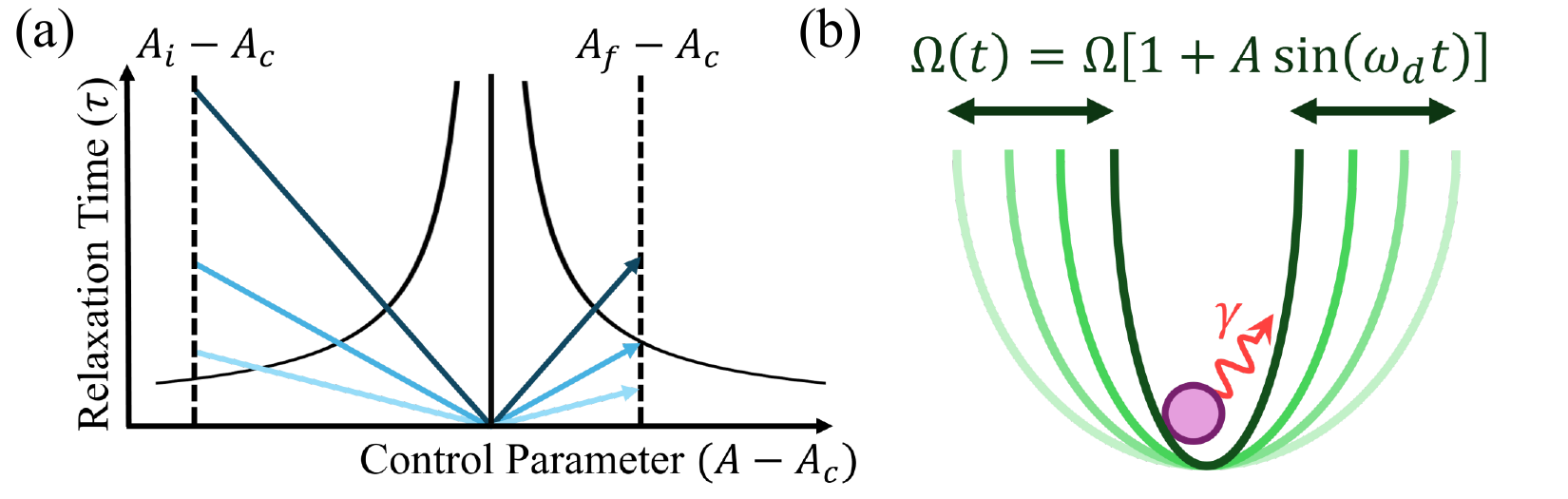}
    \caption{(a) Sketch of the adiabatic-impulse approximation for finite ramp protocols. (b) Sketch of a generic dissipative linear parametric oscillator.}
    \label{fig:schematics}
\end{figure}

We begin with a brief review of the AI approximation and how it leads to the KZM. Consider a generic system that can undergo a continuous phase transition from a disordered to an ordered phase by tuning a control parameter $A$ across some critical point $A_{c}$. Near the critical point, the system's relaxation time $\tau$ and correlation length $\xi$ diverges, with $\tau \propto |\varepsilon|^{-vz}$ and $\xi \propto |\varepsilon|^{-v}$, where $\varepsilon = \left( A - A_{c} \right) / A_{c}$ \cite{del_campo_universality_2014}. Here, $v$ and $z$ are the static and dynamical critical exponents, respectively. Suppose we linearly ramp $A$ using the protocol,
\begin{equation}
\label{eq:quench_protocol}
A(t) = 
\begin{cases}
A_{i}\quad \quad \quad \quad \quad \quad \quad \quad \quad t_{i} \leq t \leq 0 \\
\left(A_{f} - A_{i}\right) \left(\frac{t}{\tau_{q}} \right) + A_{i}, \quad 0 < t < \tau_{q}, \\
A_{f} \quad \quad \quad \quad \quad \quad \quad \quad \quad \tau_{q} \leq t \leq t_{f}, 
\end{cases}, 
\end{equation}
where $A_{i}$ ($A_{f}$) is the initial (final) value of the control parameter, and $t_{i}$ ($t_{f}$) is the initial (final) time. We initialize the system deep in the disordered phase, $A_{i} \ll A_{c}$, and assume that $\tau_{q}$ is sufficiently large. Initially, the system remains in an adiabatic regime where all of its macroscopic quantities adiabatically follow the quench. As $A$ approaches $A_{c}$, the system enters an impulse regime, in which the divergence of $\tau$ causes the system to remain frozen in the disordered phase even after the system has passed $A_{c}$ at the critical time $t_{c}$. The system only leaves the impulse regime and enters a new phase after some transition time $t_{p}$, with $t_{p} > t_{c}$. This crossover of the system from an adiabatic dynamics to a frozen dynamics constitutes the AI approximation of the KZM \cite{del_campo_universality_2014}. A sketch of this approximation is shown in Fig.~\ref{fig:schematics}(a).

One of the key assumptions of the KZM is that the relaxation time at $A(t_{p})$ sets the transition delay $\hat{t} \equiv t_{p} - t_{c}$, with $\hat{t} \propto \tau\left(A(t_{p}) \right)$ \cite{del_campo_universality_2014}. Meanwhile, the correlation length at $t_{p}$, $\xi(t_{p})$, sets the number of spatial defects $n_{d}$ post-quench. Specifically, $n_{d}$ follows the relation 
\begin{equation}
\label{eq:defect_correlation_relations}
n_{d} \propto \xi(t_{p})^{-(D - d_{f})},
\end{equation}
where $D$ is the system's dimension, and $d_{f}$ is the defect dimension \cite{del_campo_universality_2014}. These two assumptions lead to the main signatures of the KZM: the power-law scaling of $\hat{t}$, $\xi(t_{p})$, and $n_{d}$ with $\tau_{q}$, 
\begin{equation}
\label{eq:kzm_scaling}
\hat{t} \propto \tau_{q}^{\frac{vz}{1 + vz}}, \quad \xi(t_{p}) \propto \tau_{q}^{\frac{v}{1 + vz}}, \quad n_{d} \propto \tau_{q}^{-(D-d_{f}) \frac{v}{1 + vz}}.
\end{equation}
Showing that a quenched system exhibits these scaling behaviors is enough to demonstrate that the system follows the KZM \cite{del_campo_universality_2014}. Note that for a finite ramp protocol, the KZM is expected to break down at very small values of $\tau_{q}$, resulting in a constant $\hat{t}$, $\xi(t_{p})$, and $n_{d}$ \cite{zeng_universal_2023, jara_jr_apparent_2024}.

Now, to show that an open DTC forming system can exhibit signatures of the KZM, we have to first verify that it satisfies the condition for the AI approximation. This condition is the divergence of the relaxation time near the critical point. To do this, consider the DLPO, which is the simplest system that exhibits a period-doubling response. This system is described by the equation
\begin{equation}
\label{eq:dlpo_eom}
\ddot{\theta} + \gamma \dot{\theta}+ \Omega^{2}\left[ 1 + A \sin(\omega_{d}t) \right]\theta = 0,
\end{equation} 
where $\gamma$ is the dissipation rate, $\Omega$ is the natural frequency, and $\omega_{d}$ and $A$ are the driving frequency and amplitude, respectively. A sketch of this system is shown in Fig.~\ref{fig:schematics}(b). As shown in Ref.~\cite{jara_jr_theory_2024}, a huge class of systems that can form DTCs can be mapped onto a DLPO. In the resonance limit of $\omega_{d} = 2\Omega$ and $A \ll 1$, we can obtain an approximate analytic solution for Eq.~\eqref{eq:dlpo_eom}, which takes the form \cite{kovacic_mathieus_2018} (see Appendix~\ref{sec:dlpo_solution} for details), 
\begin{equation}
\label{eq:dlpo_full_solution}
\theta(t) = \left( \alpha_{0} e^{-\omega_{d}t / \tau_{-}} + i \beta_{0}e^{-\omega_{d}t / \tau_{+}}  \right)e^{i\omega_{d}t/ 2} + \mathrm{cc}., 
\end{equation}
where $\alpha_{0}$ and $\beta_{0}$ are constants set by the initial conditions of the DLPO, and $\tau_{\pm}$ are the characteristic relaxation timescales
\begin{equation}
\tau_{\pm} = 8 (A_{c} \pm A)^{-1}, \quad A_{c} = 2\gamma / \Omega.
\end{equation}
Below the critical point $A_{c}$, the DLPO is in a disordered state, in which, due to positive $\tau_{\pm}$, any oscillation with large amplitude exponentially decays until the dynamics becomes dominated by noise, if present. Above $A_{c}$, $\tau_{-}$ becomes negative, resulting in an exponential growth of $\theta$ and the emergence of a resonant state with period-doubling response.

Near $A_{c}$ for $A < A_{c}$, we can observe that as $A\rightarrow A_{c}$, $\tau_{-}$ diverges while $\tau_{+}$ remains finite. As a result, the dynamics of the DLPO can be effectively described by the equation,
\begin{equation}
\theta \approx \alpha_{0}e^{-\omega_{d}t / \tau_{-}}\cos\left( \omega_{d}t/ 2 \right).
\end{equation}
Having identified $\tau_{-}$ as the main relaxation time scale $\tau$ near the critical point, we can now see that the DLPO satisfy the criterion for the AI approximation, since $\tau$ algebraically diverges as $A \rightarrow A_{c}$ with an exponent of $vz = 1$. Following this observation, we can infer that if a system with dimension $D \geq 1$ can be mapped onto a DLPO, then the AI approximation must hold for the system when quenched to form a DTC. Therefore, we should be able to observe the main signatures of the KZM. It also predicts that for any DTC-forming system that can be mapped onto or approximated as a DLPO, its relaxation time should diverge as $\tau \propto |A - A_{c}|^{-vz} \propto |A-A_{c}|^{-1}$, implying a universality class characterized by the critical exponent, $vz =1$.

\section{Systems of Interest}\label{sec:system}

\begin{figure}
    \centering
    \includegraphics[scale = 0.4]{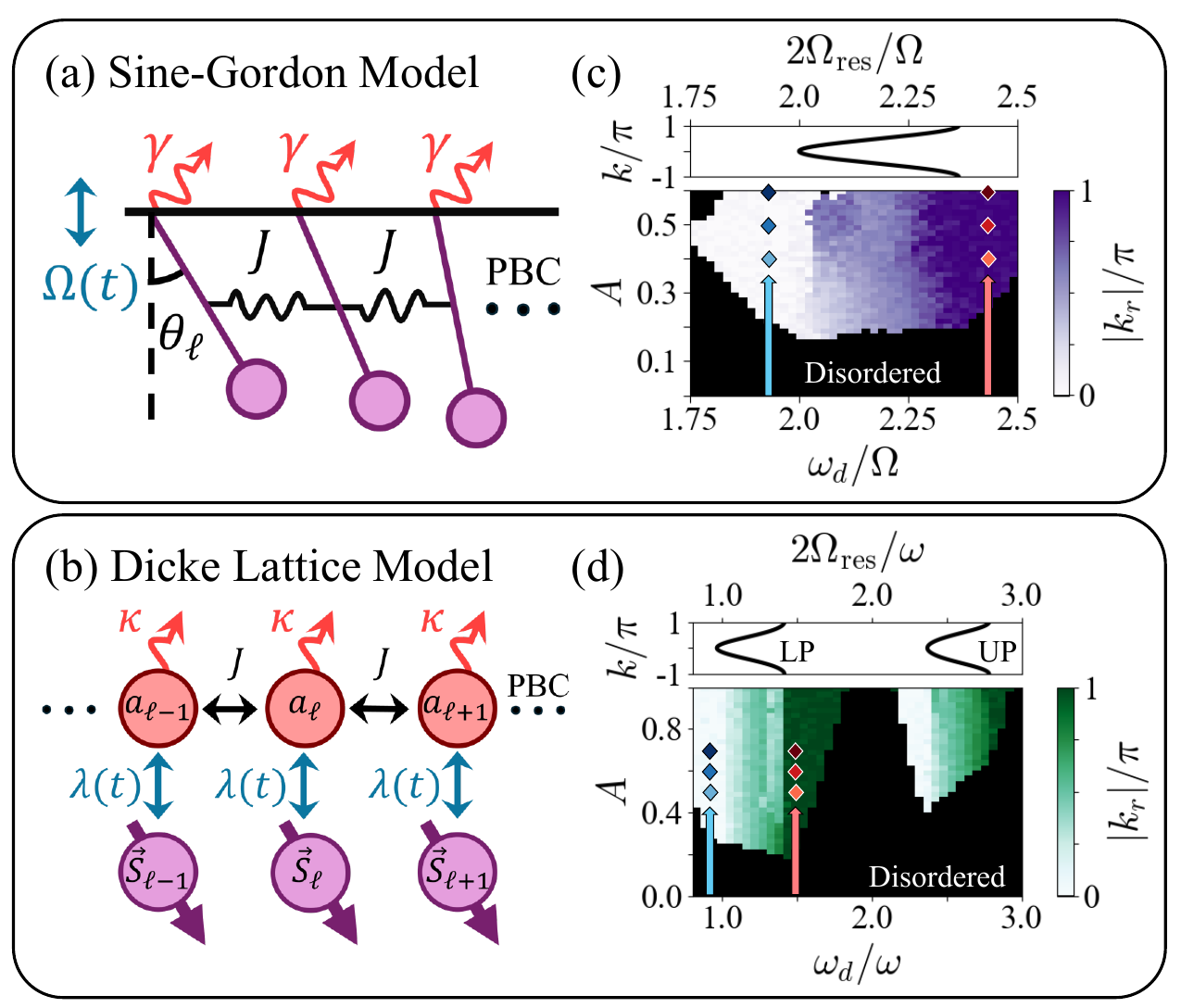}
    \caption{(a)--(b) Sketch of the (a) Sine-Gordon model and (b) open Dicke lattice model. (c)--(d) Phase diagram of (c) the SGM and (d) the DLM. Top panels show the frequency associated with the dispersion relation of the two systems, $\Omega_{\mathrm{res}}(k)$, while the bottom panels show the corresponding phase diagrams for the two systems. The parameters considered for the SGM are $\left\{ \gamma, g, \tilde{T}, M \right\} = \left\{ 0.1\Omega, 0.1 \Omega^{2}, 10^{-5}, 100 \right\}$, while we consider the parameters $\left\{ \lambda_{0}, \kappa, J, N, M \right\} = \left\{ 0.7 \lambda_{c}, 0.1 \omega, 0.1 \omega, 10^{4}, 100 \right\}$ for the DLM. The arrows in the bottom panels depict the direction of the linear ramp considered in Fig.~\ref{fig:kzm_scaling}, while the diamond markers mark the considered values of $A_{f}$ in Fig.~\ref{fig:kzm_scaling}.}
    \label{fig:phase_diagram}
\end{figure}

We now test the validity of the KZM in DTC-forming systems by investigating exemplary one-dimensional systems with periodic boundaries. We first consider the SGM, which is an array of $M$ coupled classical pendula as shown in Fig.~\ref{fig:phase_diagram}(a). It is described by the equation
\begin{equation}
\label{eq:sg_model_eom}
\ddot{\theta}_{\ell} + \gamma \dot{\theta}_{\ell} + f(t)\sin\theta_{\ell} - g\left( \theta_{\ell - 1} + \theta_{\ell + 1} - 2\theta_{\ell} \right) = \eta_{\ell},
\end{equation}
where $g$ is the nearest-neighbor coupling strength, and $f(t)=\Omega^{2} \left[1 + A(t)\sin(\omega_{d}t) \right]$ is the periodic driving. The Gaussian white noise $\eta_{\ell}$ represents the thermal bath connected to each pendulum. It satisfies the relation $\langle \eta_{\ell}(t) \rangle = 0$ and $\langle \eta_{\ell}(t)\eta_{\ell'}(t') \rangle = 2\tilde{T}\Omega^{2}\gamma\delta(t - t')\delta_{\ell, \ell'}$, where $\tilde{T} = k_{B}T / \left( mL^{2}\Omega^{2} \right)$ is the unitless temperature \cite{yao_classical_2020}. Here, $m$ and $L$ are the characteristic mass and length of each pendulum, respectively, and $k_{B}$ is the Boltzmann constant. In the context of phase transitions and non-equilibrium physics, the SGM and its variants are known to follow the KZM in the static limit \cite{laguna_critical_1998, laguna_density_1997, suzuki_deconstructing_2025, zeng_universal_2023}. This makes it an appropriate toy model for determining whether classical periodically-driven systems also follow the KZM.

The second system we consider is the DLM, described by the Lindblad master equation \cite{zou_implementation_2014} 
\begin{equation} 
\label{eq:dlm_master_eq}
\partial_{t} \hat{\rho} = -i \left[\hat{H}/\hbar, \hat{\rho} \right] + \kappa \sum_{\ell = 1}^{M} \left( 2\hat{a}_{\ell} \hat{\rho}\hat{a}_{\ell}^{\dagger} - \left\{ \hat{a}_{\ell}^{\dagger}\hat{a}_{\ell}, \hat{\rho} \right\} \right), 
\end{equation}
where the Hamiltonian reads
\begin{equation}
\frac{\hat{H}}{\hbar} = \sum_{\ell = 1}^{M} \frac{\hat{H}_{\ell}}{\hbar} - J\left( \hat{a}_{\ell}^{\dagger}\hat{a}_{\ell + 1} + \hat{a}_{\ell + 1}^{\dagger}\hat{a}_{\ell} \right),
\end{equation}
with $\hat{H}_{\ell}$ being the Dicke Hamiltonian in each site,
\begin{equation}
\frac{\hat{H}_{\ell}}{\hbar} =  \omega \hat{a}_{\ell}^{\dagger}\hat{a}_{\ell} + \omega_{0} \hat{S}^{z}_{\ell} + \frac{2\lambda}{\sqrt{N}}\left(\hat{a}_{\ell}^{\dagger} + \hat{a}_{\ell} \right)\hat{S}^{x}_{\ell} . 
\end{equation}
The DLM, schematically depicted in Fig.~\ref{fig:phase_diagram}(b), describes the dynamics of an array of $M$ coupled spin-cavity systems. Each site is composed of $N$ spin-$1/2$ particles, or qubits, coupled to a single lossy bosonic mode, represented by the annihilation operator $\hat{a}_{\ell}$. The qubits in each site are represented by the collective spin operator $\hat{S}^{x, y, z}_{\ell}$. Here, $\omega$ and $\omega_{0}$ are the boson and spin transition frequencies, respectively, $\lambda(t) = \lambda_{0}\left[ 1 + A(t)\sin\left( \omega_{d}t \right) \right]$ is the time-dependent spin-boson coupling, and $J$ is the nearest-neighbour coupling between the bosonic modes. In the static limit, $A=0$, the DLM has two phases. Below the critical spin-boson coupling
\begin{equation}
\lambda_{c} = \frac{1}{2}\sqrt{\omega_{0} \left[ \left(\omega - 2J \right)^{2} + \kappa^{2} \right] / (\omega - 2J)}
\end{equation}
for $J \ll \omega, \omega_{0}$, the system is in a disordered normal phase, in which the bosonic modes are in the vacuum state, and the collective spin operators are polarized in the $-z$ direction. Above $\lambda_{c}$, the system enters a homogeneous superradiant phase, in which the bosonic modes attain macroscopic excitations, while the collective spins get a nonzero $S_{x} \equiv \langle S_{x} \rangle$ component, with a sign randomly chosen from its two degenerate states. As shown in Ref.~\cite{jara_jr_apparent_2024}, similar to the SGM, the transition of the DLM from a disordered phase to the homogeneous superradiant phase through a linear ramp follows the KZM. This makes the DLM a suitable model for testing the KZM in periodically driven quantum systems. Throughout this work, we set $\omega = \omega_{0}$, and restrict our attention to $\lambda_{0} < \lambda_{c}$, since in this regime, the DLM can be mapped onto a DLPO \cite{jara_jr_apparent_2024, zou_implementation_2014}. We also employ the truncated Wigner approximation to obtain the system's dynamics. This approximation allows us to incorporate the first-order correction to the quantum fluctuations without solving the dynamics of the system's density matrix \cite{polkovnikov_phase_2010}. We provide more details on the approximation in Appendix \ref{sec:dlm_twa}.

We present in Figs.~\ref{fig:phase_diagram}(c) and \ref{fig:phase_diagram}(d) the phase diagram of the SGM and DLM, respectively, spanned by $\omega_{d}$ and a fixed driving amplitude, $A(t) = A$. The black regions correspond to the disordered phase, in which fluctuations dominate the dynamics of the two systems, while the colored regions mark the parameter regimes where period-doubled DTCs can form. To identify whether a system is in a period-doubled DTC, we consider a local order parameter $O_{\ell}(t)$ exhibiting period-doubling oscillations in the DTC phase. The complex amplitude \cite{apffel_experimental_2024, jara_controlled_2025}
\begin{equation}
O_{\ell, R}(t) = \frac{\omega_{R}}{\pi} \int_{t}^{t + 2\pi / \omega_{R}} O_{\ell}(t')e^{i\omega_{R}t'}dt' ,
\end{equation}
with $\omega_{R} = \omega_{d}/2$, captures the response amplitude, $|O_{\ell, R}|$, and the modulation phase shift, $\varphi_{\ell} \equiv \arg\left\{ O_{\ell, R} \right\}$, of $O_{\ell}$. The system is in a period-doubled DTC if $|O_{\ell, R}| > 0$ and $\varphi_{\ell}$ reaches a steady state; otherwise, it is in a disordered phase. We find that in the weak noise limit, which corresponds to $\tilde{T} \ll 1$ and $N \gg 1$ for the SGM and DLM, respectively, DTCs can form by driving the two systems near twice the frequency associated with their respective dispersion relations, $\Omega_{\mathrm{res}}(k)$. Depending on the chosen $\omega_{d}$, the emerging DTC can have modulations in space, characterized by the wave vector $|k_{r}|$. In particular, if the two systems are driven at $\omega_{d} \leq 2\Omega_{\mathrm{res}}(k=0)$, the resulting DTC is a ferromagnetic DTC (FM-DTC), in which all sites have the same $\varphi_{\ell}$, and $|k_{r}| = 0$. Meanwhile, if $\omega_{d} \geq 2\Omega_{\mathrm{res}}(k=\pi)$, both systems form an antiferromagnetic DTC (AFM-DTC) in which $\varphi_{\ell}$ is shifted by $\pi$ at every site, and $|k_{r}| = \pi$. As shown in Figs.~\ref{fig:phase_diagram}(c) and \ref{fig:phase_diagram}(d), for both systems, the transition from FM-DTC to AFM-DTC is continuous, evinced by the smooth change in the color gradient associated with $|k_{r}|$. This tunability of $|k_{r}|$ allows us to test the validity of the KZM for arbitrary spatiotemporal order.

\section{Kibble-Zurek mechanism for discrete time crystals}\label{sec:kzm_dtc}

\begin{figure}
    \centering
    \includegraphics[scale = 0.35]{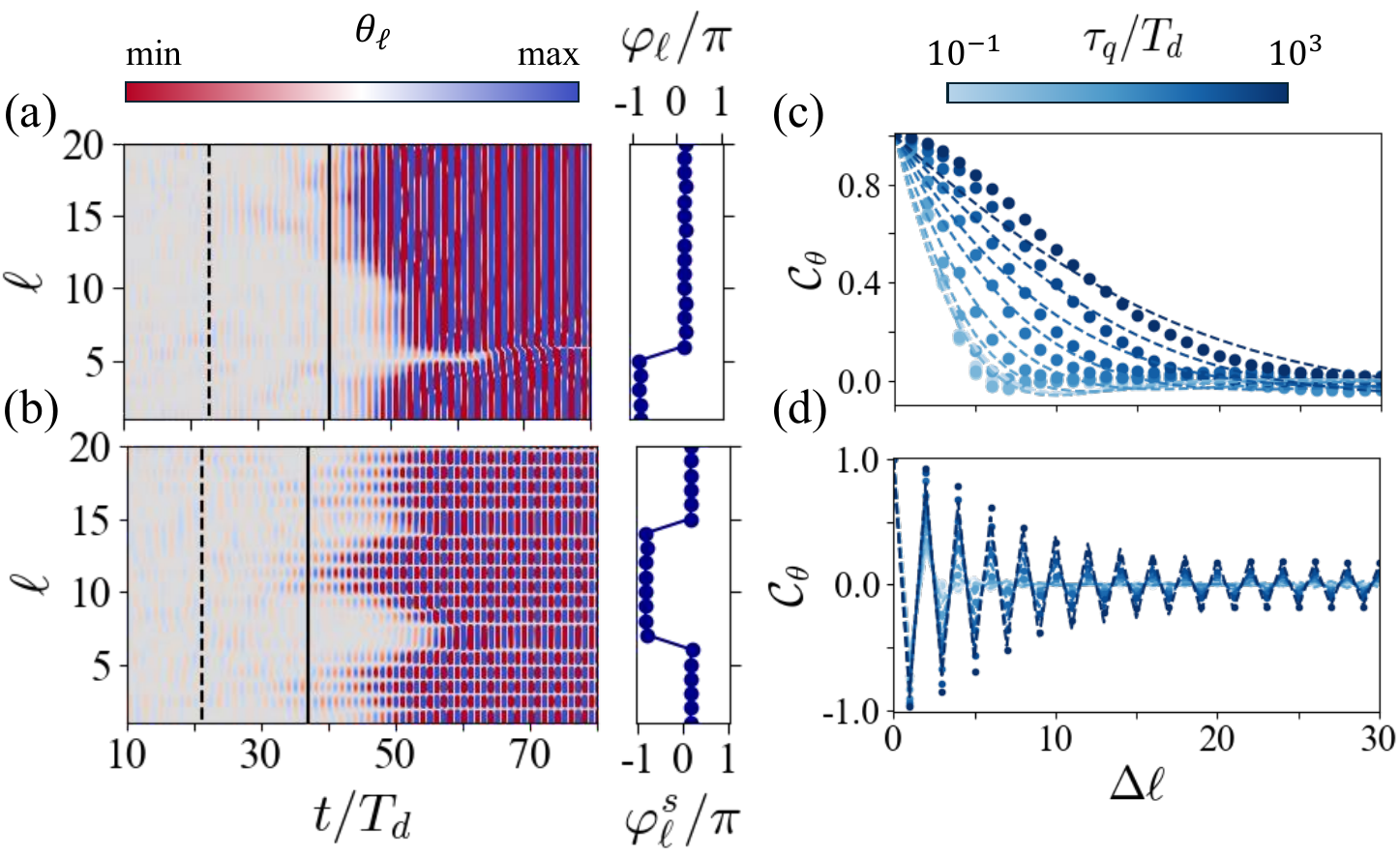}
    \caption{(a)--(b) Left panels show the exemplary spatiotemporal dynamics of the SGM for (a) the FM-DTC and (b) the AFM-DTC and for $\tau_{q} = 50 T_{d}$. The solid lines correspond to the transition time $t_{p}$, while the dashed lines represent the critical time $t_{c}$. Right panels show the steady state profile of the modulation phase shift $\varphi_{\ell}$ and its staggered counterpart $\varphi_{\ell}^{s}$ obtained at the final time $t_{f}$. (c)--(d) Spatial correlation of the SGM at $t_{p}$ for (c) the FM-DTC and (d) the AFM-DTC, for different values of $\tau_{q}$. The dashed lines represent the best fit curve associated with the numerically obtained values of $C_{\theta}$. The parameters considered are $\left\{ \gamma, g, \tilde{T}, M  \right\} = \left\{ 0.1\Omega,  0.1 \Omega^{2}, 10^{-5}, 100 \right\}$, with $\omega_{d} = 1.9\Omega$ for the FM-DTC and $\omega_{d} = 2.4\Omega$ for the AFM-DTC.}
    \label{fig:sgm_dynamics}
\end{figure}

We now provide a recipe for testing the KZM in any lattice system transitioning into a DTC by increasing the driving amplitude via the ramp in Eq.~\eqref{eq:quench_protocol}. Consider the transition dynamics of the SGM from a disordered phase into either an FM-DTC or AFM-DTC, as shown in Figs.~\ref{fig:sgm_dynamics}(a) and \ref{fig:sgm_dynamics}(b), respectively. Note that the DTC formation in the DLM follows the same behavior as the SGM, as presented in Appendix \ref{sec:dlm_twa}. We determine $t_{p}$ using a threshold criterion, in which we consider the time the system begins forming DTCs to be
\begin{equation}
\sum_{\ell = 1}^{M}|O_{\ell, R}(t = t_{p})| = \delta \max_{t} \left\{\sum_{\ell = 1}^{M}|O_{\ell, R}(t)|\right\}
\end{equation}
where $\delta = 0.15$ is an arbitrary threshold. Here, $\max_{t}\left\{ \sum_{\ell} |O_{\ell, R}(t)| \right\}$ is proportional to the steady state value of $|O_{\ell, R}|$, which is independent of the quench time. For $t_{c}$, we use the relation 
\begin{equation}
\label{eq:critical_time}
t_{c} = \left( A_{c} - A_{i} \right)\tau_{q} / \left( A_{f} - A_{i} \right) 
\end{equation}
inferred from Eq.~\eqref{eq:quench_protocol}. The $A_{c}$ is identified by considering the scaling of $A(t_{p})$ with $\tau_{q}$ and extrapolating its value as $\tau_{q} \rightarrow \infty$ (see Appendix \ref{sec:critical_point} for details). In Figs.~\ref{fig:sgm_dynamics}(a) and \ref{fig:sgm_dynamics}(b), we present the $t_{p}$ and $t_{c}$ for the two DTC configurations as solid and dashed lines, respectively. The difference between these quantities gives the transition delay $\hat{t}$.

\begin{figure}
    \centering
    \includegraphics[scale = 0.36]{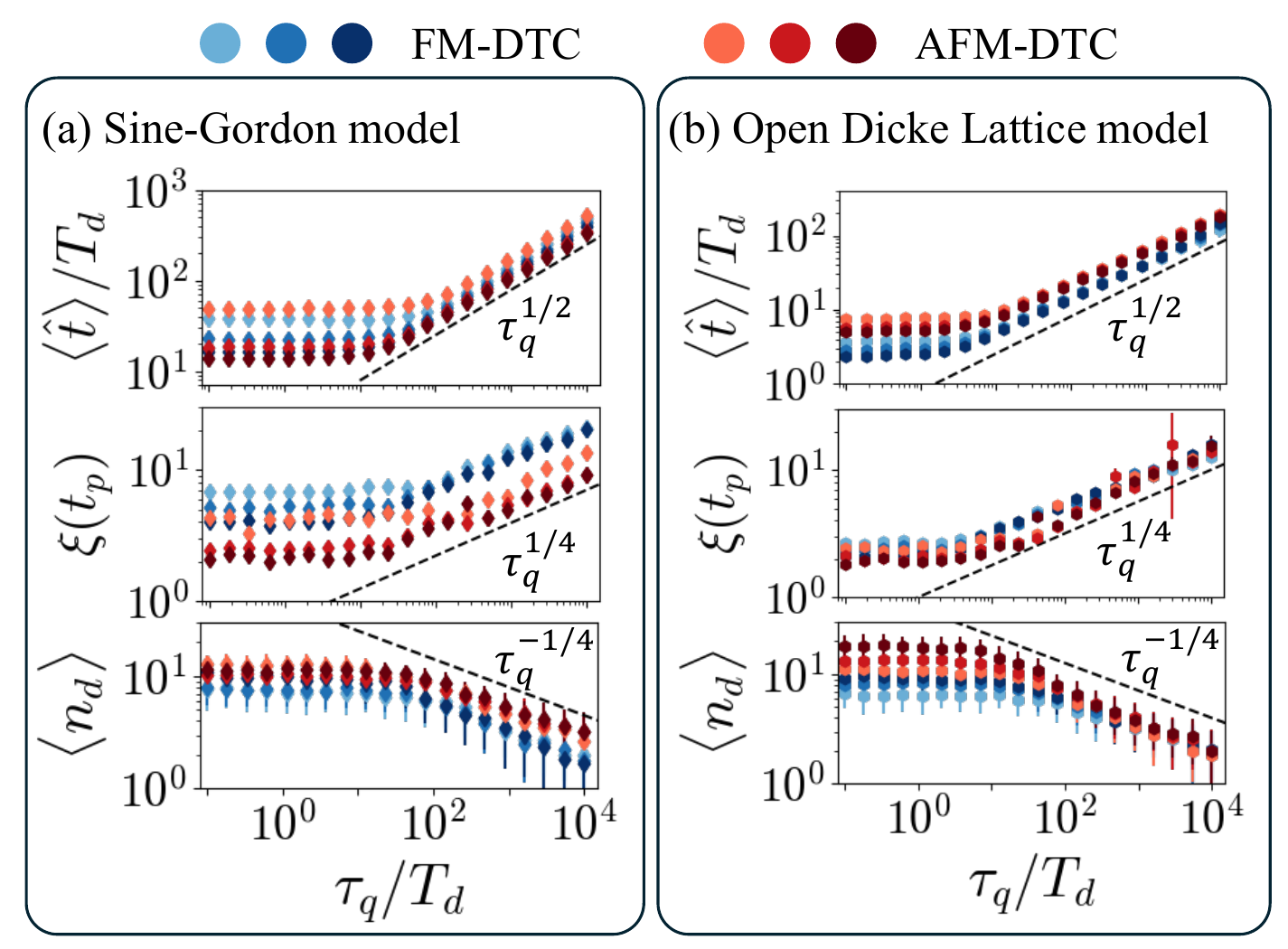}
    \caption{(a)--(b) KZM scaling of (a) the SGM and (b) the DLM for different DTC configurations and of $A_{f}$. The parameters considered for the SGM are $\left\{ \gamma, g, \tilde{T}, M \right\} = \left\{ 0.1\Omega, 0.1\Omega^{2}, 10^{-5}, 100 \right\}$, with $\omega_{d} = 1.9\Omega$ for the FM-DTC and $\omega_{d} = 2.4\Omega$ for the AFM-DTC. For the SGM, $A_{f} \in \left\{ 0.4, 0.5, 0.6 \right\}$. As for the DLM, we consider the parameters $\left\{ \lambda_{0}, \kappa, J, N, M  \right\} = \left\{ 0.7\lambda_{c}, 0.1\omega, 0.1\omega, 10^{4}, 100 \right\}$, with $\omega_{d} = 0.9\omega$ for the FM-DTCs and $\omega_{d} = 1.5\omega$ for the AFM-DTCs. The $A_{f}$ for the DLM are $A_{f} \in \left\{ 0.5, 0.6, 0.7 \right\}$. Dashed lines correspond to the power-law scaling predicted for $v = 1/2$, $z = 2$. Increasing saturation of data points corresponds to increasing values of $A_{f}$.}
    \label{fig:kzm_scaling}
\end{figure}

We next determine the steady state $n_{d}$ by looking at the steady state spatial profile of $\varphi_{\ell}$ obtained at the final time $t_{f}$. As shown in Fig.~\ref{fig:sgm_dynamics}(a), for the FM-DTC, the defect in the weak noise limit manifests as a phase slip, in which two domains are $\pi$-shifted with one another. As for the AFM-DTC shown in Fig.~\ref{fig:sgm_dynamics}(b), the phase slip manifests instead in the staggered modulation phase,
\begin{equation}
\varphi_{\ell}^{s} = \varphi_{\ell} + \left[ 1 - (-1)^{\ell} \right] \pi / 2.
\end{equation}
We note that non-trivial spatio-temporal defects may also appear in the strong noise limit, $\tilde{T} < 1$ and $N > 1$. They are induced by fluctuations, however, and thus vanish in the weak-noise limit, as we have demonstrated in Appendix~\ref{sec:spacetime_defect}. We identify the location of the defect in the FM-DTC (AFM-DTC) by calculating the nearest neighbour difference $\Delta \varphi_{\ell}^{(s)} = \varphi_{\ell + 1}^{(s)} - \varphi_{\ell}^{(s)}$, taking note of the periodic boundary condition of the system. We consider two neighbouring sites to be in the same domain if $|\Delta \varphi_{\ell}^{(s)}| < \pi/2$. Otherwise, a phase slip occurs between sites $\ell$ and $\ell + 1$. As shown in Figs.~\ref{fig:sgm_dynamics}(a) and \ref{fig:sgm_dynamics}(b), and in Figs.~\ref{fig:dlm_dynamics}(a) and \ref{fig:dlm_dynamics}(b) for the DLM, the set threshold is adequate to differentiate distinct domains in the system.

Finally, to determine the correlation length at $t_{p}$, $\xi(t_{p})$, we obtain the equal-time spatial correlation 
\begin{equation}
\mathcal{C}_{O}(\Delta \ell) = \frac{\sum_{\ell = 1}^{M} \mathrm{Re} \left[ \left< O^{*}_{\ell + \Delta \ell, R}(t_{p})O_{\ell, R}(t_{p}) \right> \right]}{\sum_{\ell=1}^{M} |O_{\ell, R}(t_{p})|^{2}},
\end{equation}
where $\langle \cdots \rangle$ corresponds to an ensemble average. We present in Fig.~\ref{fig:sgm_dynamics}(c) and \ref{fig:sgm_dynamics}(d) the spatial correlation of the SGM for the FM-DTC and the AFM-DTC, respectively. For the FM-DTC, the spatial correlation exponentially decays with the separation distance $\Delta \ell$. Meanwhile, the spatial correlation for the AFM-DTC exhibits oscillations that exponentially decay at a finite value. Note that we observe the same behavior for the spatial correlations in the DLM, as shown in Appendix~\ref{sec:dlm_twa}. For both DTC configurations, we quantify the correlation length by fitting the function
\begin{equation}
\mathcal{C}_{O, \mathrm{fit}} = \left[ e^{-\ell/\xi(t_{p})} + c_{0} \left( 1 - e^{-\ell / \xi(t_{p})} \right)\right]\cos(k_{r}\ell)
\end{equation}
on the numerically obtained $\mathcal{C}_{O}$. In this function, the parameter $c_{0}$ captures the saturation of $\mathcal{C}_{O}$ at a finite value, while $k_{r}$ captures the wave number of the DTC.

We now present in Figs.~\ref{fig:kzm_scaling}(a) and \ref{fig:kzm_scaling}(b) the scaling of $\langle \hat{t} \rangle$, $\xi(t_{p})$, and $\langle n_{d} \rangle$ as a function of $\tau_{q}$ for the SGM and the DLM, respectively, averaged over $100$ noise realizations. We set $O_{\ell} = \theta_{\ell}$ to be the main order parameter for the SGM, while we set $O_{\ell} = a_{\ell}$ for the DLM. We can observe that for large $\tau_{q}$, $\langle \hat{t} \rangle$, $\xi(t_{p})$, and $\langle n_{d} \rangle$ all follow a power-law scaling independent of $\omega_{d}$ and $A_{f}$, suggesting the validity of the KZM at large $\tau_{q}$. For reference, we also include in Figs.~\ref{fig:kzm_scaling}(a) and \ref{fig:kzm_scaling}(b) the expected scaling of $\hat{t}$, $\xi(t_{p})$, and $n_{d}$ for the mean-field Ising universality class, $v = 1/2$ and $z = 2$ \cite{del_campo_universality_2014}. For small $\tau_{q}$, all three quantities saturate at a finite value, signalling the breakdown of the KZM at rapid quenches \cite{zeng_universal_2023}. This power-law behavior of $\langle \hat{t} \rangle$, $\xi(t_{p})$, and $\langle n_{d} \rangle$ demonstrates the validity of KZM in phase transitions involving genuine dynamical order, such as DTCs. Note that finite-size effects are negligible for the system sizes considered here, as evinced by $\xi(t_{p})$ being an order of magnitude smaller than $M$ for all $\tau_{q}$ considered.

\section{Universality of discrete time crystal formation}\label{sec:universality}

\begin{figure}
    \centering
    \includegraphics[scale = 0.62]{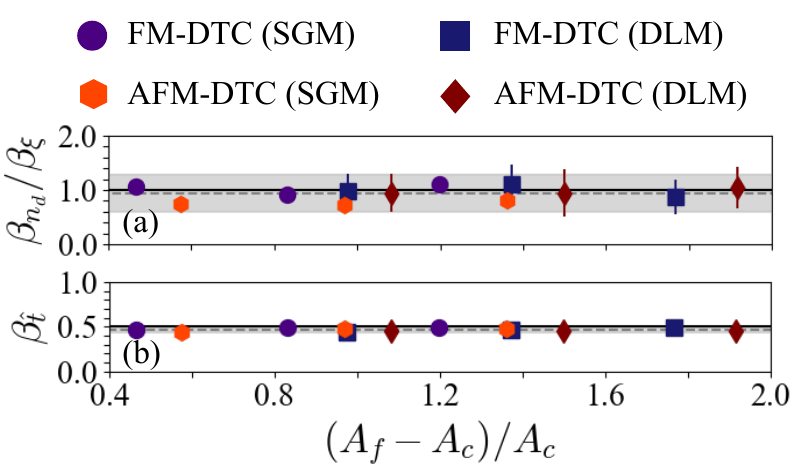}
    \caption{(a) Ratio of the scaling exponents of $\xi(t_{p})$ and $\langle n_{d} \rangle$ for the SGM and the DLM. (b) Scaling exponent of $\langle \hat{t} \rangle$ for both the DLM and the SGM. The solid line in (a) corresponds to the theoretical ratio $\beta_{n_{d}}/\beta_{\xi} = 1$ for one-dimensional systems with point defects, while in (b), it corresponds to the theoretical $vz$ for the DLPO. The dashed lines represent the mean value of $\beta_{n_{d}}/\beta_{\xi}$ and $\beta_{\hat{t}}$ averaged over different values of $A_{f}$, while the gray regions mark the uncertainty of the estimated mean values. The scaling exponents shown here are extracted from the best curve fit of the data points in Fig.~\ref{fig:kzm_scaling}.}
    \label{fig:scaling_exponent}
\end{figure}

Given the validity of the KZM in DTC-forming systems, we now address whether we can identify their universality class by using the KZM. To this end, we first establish the relationship between $\xi(t_{p})$ and $\langle n_{d} \rangle$. As discussed, the KZM predicts that $\xi(t_{p})$ sets the final number of defects in the system after the quench. If this prediction holds, given that $\xi(t_{p}) \propto \tau_{q}^{\beta_{\xi}}$ and $ \langle n_{d} \rangle \propto \tau_{q}^{-\beta_{n_{d}}}$, then we should obtain $\beta_{n_{d}} / \beta_{\xi} = D-d_{f}$. We demonstrate in Fig.~\ref{fig:scaling_exponent}(a) that for both the SGM and the DLM, independent of $A_{f}$ and the DTC configuration, $\beta_{n_{d}} / \beta_{\xi} = 1$, which is consistent for $D=1$ dimensional systems with point defects ($d_{f} = 0$). This result demonstrates that the defect formation in both systems is indeed due to the KZM. We can also note from this result that the critical exponents can be obtained by only considering either $\beta_{\xi}$ or $\beta_{n_{d}}$, and the scaling exponent of $\langle \hat{t} \rangle$.

We now establish the universality class of the SGM and the DLM by considering the scaling exponent of $\langle \hat{t} \rangle$, $\beta_{\hat{t}} = vz / (1 + vz)$. As we have predicted, any DTC-forming system that can be mapped onto a DLPO should have a $vz = 1$, which translates to $\beta_{\hat{t}} = 1/2$. As we demonstrate in Fig.~\ref{fig:scaling_exponent}(b), both systems indeed have $\beta_{\hat{t}} = 1/2$ independent of $A_{f}$ and the final DTC configuration. This result confirms that they share the same universality class as the DLPO. Therefore, we have verified that the notion of universality class extends to transitions involving dynamical phases, such as DTCs.

\begin{figure}
    \centering
    \includegraphics[scale = 0.62]{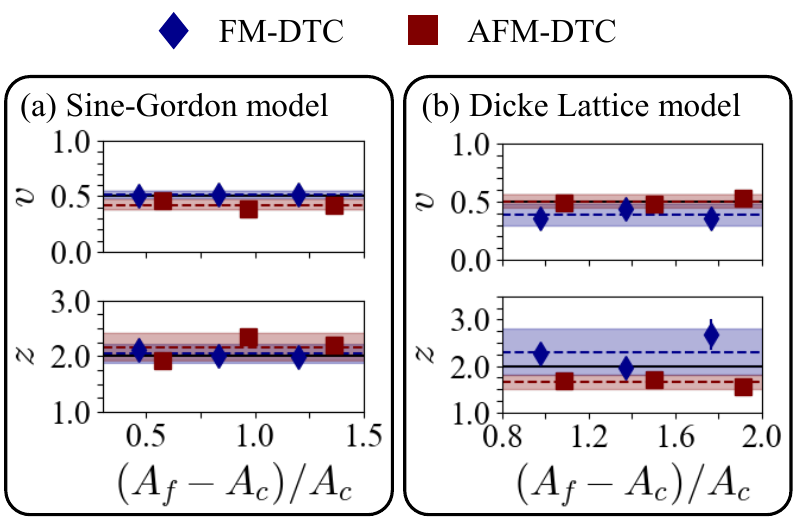}
    \caption{(a)--(b) Extracted $v$ and $z$ for (c) the SGM and the (d) DLM. The solid lines indicate the critical exponents for the mean-field Ising universality class. For all figures, the dashed lines correspond to the average of the quantity considered, while the shaded regions indicate the standard deviation. The scaling exponents shown here are extracted from the best curve fit of the data points in Fig.~\ref{fig:kzm_scaling}. }
    \label{fig:vz_exponents}
\end{figure}

For completeness, we also calculate the static and dynamical critical exponents, $v$ and $z$ respectively, for the two systems using the relations
\begin{equation}
v = \frac{\beta_{n_{d}}}{1 - \beta_{\hat{t}}}, \quad z = \frac{\beta_{\hat{t}}}{\beta_{n_{d}}}.
\end{equation}
The obtained values of $v$ and $z$ for the SGM and the DLM are shown in Figs.~\ref{fig:vz_exponents}(a) and \ref{fig:vz_exponents}(b), respectively, for different values of $A_{f}$. Notably, while both systems share the same $vz$, their individual static and dynamical critical exponents differ from one another. In particular, for the SGM, the obtained critical exponents are $v_{\mathrm{FM}} = 0.5(2)$ and $z_{\mathrm{FM}} = 2.0(2)$ for the FM-DTC, and $v_{\mathrm{AFM}} = 0.4(2)$ and $z_{\mathrm{AFM}} = 2.2(2)$ for the AFM-DTC. These are all consistent with the critical exponents of the mean-field Ising universality class up to within the uncertainty \cite{del_campo_universality_2014}. Note that these values are obtained by averaging the values of $v$ and $z$ for different $A_{f}$. As for the DLM, we have $v_{\mathrm{FM}} = 0.38(9)$ and $z_{\mathrm{FM}} = 2.3(5)$ for the FM-DTC, and $v_{\mathrm{AFM}}=0.50(6)$ and $z_{\mathrm{AFM}} = 1.6(2)$ for the AFM-DTC, which all deviate from the critical exponents of the mean-field Ising universality class. Nevertheless, the obtained $vz$ for the two configurations of the DLM are $vz_{\mathrm{FM}} = 0.8(5)$ and $vz_{\mathrm{AFM}} = 0.8(2)$, which are still consistent with the predicted $vz$ for systems falling under the universality class of the DLPO.

\section{Summary and Discussion}\label{sec:conclusion}

We have demonstrated the validity of the KZM in periodically-driven open systems transitioning from a disordered phase to a spatiotemporally ordered phase, both in the classical and quantum regimes. We have then identified a universality class for DTC-forming systems characterized by the product of their static and dynamical critical exponents, $vz$. Thus, our work suggests that DTC formation is a genuine phase transition, which goes beyond the nonlinear dynamics framework of DTCs and highlights the many-body nature of this dynamical phase. While our results can be tested in platforms emulating arrays of parametric oscillators \cite{heugel_classical_2019, heugel_role_2023, chen_onset_2007, mestre_network_2025, iyama_observation_2024}, and arrays of spin-boson systems \cite{zou_implementation_2014, white_cavity_2019, amsuss_cavity_2011, scigliuzzo_control_2022, astner_coherent_2017, liu_entanglement_2016}, we expect our findings to also apply to DTCs found in effectively zero-dimensional systems with all-to-all coupling, such as atoms in a single-mode cavity \cite{kesler_observation_2021}, wherein the relevant quantity becomes solely the transition time.

Given that our results here are in the regime in which the noise is sufficiently weak to preserve the coherence of the DTCs, it will be interesting to test the validity of the KZM for DTC formation in the strong noise limit. In this regime, the fluctuations can be strong enough to either seed spatio-temporal defects, such as shown in Appendix~\ref{sec:spacetime_defect}, or induce coarsening due to the movement of the spatial defects. An equally interesting extension of this work is to verify the validity of the KZM in DTC forming systems deep in the quantum regime, where phase transitions are described by the spectrum of the system's Liouvillian \cite{minganti_spectral_2018}. If the KZM for DTC forming systems do survive in the quantum regime, it will be interesting to compare the critical exponents obtained for these quantum systems with those obtained from classical systems either through scaling analysis, as done in this work, or through dimensional analysis \cite{nikoghosyan_universality_2016}. 
Finally, given that the defects in DTC-forming systems arise from the spontaneous breaking of both space and discrete time translation symmetry, it would be worthwhile to understand and rigorously classify these topological defects using methods like homotopy theory \cite{Kibble2000}.

\section*{Acknowledgements}
R.D.J. acknowledges support from the DOST-SEI under the Accelerated Science and Technology Human Resource Development Program. J.G.C. acknowledges financial support from the National Academy of Science and Technology, Philippines (NAST PHL). We also acknowledge financial support from the Department of Science and Technology (DOST) as monitored by the Philippine Council for Industry, Energy, and Emerging Technology Research and Development (DOST-PCIEERD) through Project No. 1214356. We thank P. Kongkhambut, H. Ke{\ss}ler, and A. Hemmerich for insightful discussions.

\appendix

\section{Approximate dynamics of dissipative linear parametric oscillator}\label{sec:dlpo_solution}

We obtain an approximate expression for the dynamics of the DLPO in the weak driving limit, $A \ll 1$, and weak dissipation limit, $\gamma \ll \Omega$, by performing a multi-scale analysis on Eq.~\eqref{eq:dlpo_eom}. Following Ref.~\cite{kovacic_mathieus_2018}, we consider a non-dimensional version of Eq.~\eqref{eq:dlpo_eom}, 
\begin{equation}
\label{eq:dlpo_nondimensional_eom}
\frac{d^{2}\theta}{d\tilde{t}^{2}} + A\Gamma \frac{d\theta}{d\tilde{t}} + \Delta^{2}\theta + A\Delta^{2}\sin\left( \tilde{t} \right) \theta = 0',
\end{equation}
where $\tilde{t} = \omega_{d}t$, $A\Gamma = \gamma / \omega_{d}$, $\Delta = \Omega / \omega_{d}$. In the resonance limit, $\omega_{d} = 2\Omega$ or $\Delta^{2} = 1/ 4$, we can approximate a solution for Eq.~\eqref{eq:dlpo_nondimensional_eom} by assuming that the dynamics exhibits two timescales, a fast timescale $\chi = \tilde{t}$ and a slow timescale $\zeta = A\tilde{t}$, with $\chi \gg \zeta$. In this description, the ansatz for the solution takes the form,
\begin{equation}
\label{eq:theta_ansatz}
\theta(t) \approx \theta^{(0)}(\chi, \zeta) + A\theta^{(1)}(\chi, \zeta) + A^{2}\theta^{(2)}(\chi, \zeta) + \ldots.
\end{equation}
Substituting Eq.~\eqref{eq:theta_ansatz} back to Eq.~\eqref{eq:dlpo_nondimensional_eom}, and rewriting it in terms of $\chi$ and $\zeta$ results to the equation
\begin{equation}
\label{eq:dlpo_approximate}
\left( \frac{\partial^{2} \theta^{(0)}}{\partial \chi^{2}} + \frac{1}{4}\theta^{(0)} \right) + Af + \mathcal{O}\left( A^{2} \right) + \cdots = 0,
\end{equation}
where
\begin{equation}
f =   \frac{\partial^{2} \theta^{(1)}}{\partial \chi^{2}} + \frac{1}{4}\theta^{(1)} + 2\frac{\partial^{2}\theta^{(0)}}{\partial \chi \partial\zeta} + \Gamma \frac{\partial \theta^{(0)}}{\partial \chi}  + \frac{1}{4}\sin(\chi)\theta^{(0)} .
\end{equation}
Truncating only up to $\mathcal{O}(A)$ and imposing the equality in Eq.~\eqref{eq:dlpo_approximate} results to two coupled differential equations of the form,
\begin{subequations}
\begin{equation}
\label{eq:fast_eom}
 \frac{\partial^{2} \theta^{(0)}}{\partial \chi^{2}} + \frac{1}{4}\theta^{(0)} = 0,
\end{equation}
\begin{equation}
\label{eq:slow_eom}
 \frac{\partial^{2} \theta^{(1)}}{\partial \chi^{2}} + \frac{1}{4}\theta^{(1)} + 2\frac{\partial^{2}\theta^{(0)}}{\partial \chi \partial\zeta} + \Gamma \frac{\partial \theta^{(0)}}{\partial \chi}  + \frac{1}{4}\sin(\chi)\theta^{(0)} = 0.
\end{equation}
\end{subequations}

The first equation, Eq.~\ref{eq:fast_eom}, corresponds to the fast oscillating dynamics of the DLPO. Its solution takes the form,
\begin{equation}
\label{eq:fast_eom_solution}
\theta^{(0)} = \alpha(\zeta)e^{i\chi/2} + \alpha^{*}(\zeta)e^{-i\chi/2}.
\end{equation}
The second equation, Eq.~\ref{eq:slow_eom}, meanwhile, gives the amplitude envelope of the fast oscillating dynamics. In particular, we can substitute Eq.~\eqref{eq:fast_eom_solution} back to Eq.~\eqref{eq:slow_eom} to obtain,
\begin{equation}
\label{eq:slow_eom_with_resonance}
\frac{\partial^{2}\theta^{(1)}}{\partial \chi^{2}} + \frac{1}{4}\theta^{(1)} = - ih(\zeta)e^{i\chi/2}  -\frac{i}{8}\alpha e^{i3\chi/2} + \mathrm{cc.},
\end{equation}
where  
\begin{equation}
h(\zeta) = \frac{d\alpha}{d\zeta} + \frac{\Gamma}{2} \alpha - \frac{1}{8}\alpha^{*} .
\end{equation}
Since we assume that $\theta^{(0)} \gg A\theta^{(1)}$ at all times, we suppress all resonant terms in Eq.~\eqref{eq:slow_eom_with_resonance}, which implies that $h(\zeta) = 0$. Following this condition, the equation of motion for the amplitude envelope takes the form, 
\begin{equation}
\frac{d\alpha}{d\zeta}  = - \frac{\Gamma}{2} \alpha + \frac{1}{8}\alpha^{*}.
\end{equation}
Solving for $\alpha$ and rewriting the solutions in terms of $t$ results in Eq.~\eqref{eq:dlpo_full_solution} in the main text.

\section{Semiclassical dynamics of the Dicke lattice model}\label{sec:dlm_twa}

Consider the master equation of the DLM given in Eq.~\ref{eq:dlm_master_eq}.
As mentioned, while we can compute $\hat{\rho}$ to obtain the quantum dynamics of the system, this method is computationally expensive since the Hilbert space grows exponentially with the system size. We circumvent this problem by employing the truncated Wigner approximation. We begin by approximating the equations of motion of the DLM using the Heisenberg-Langevin equation \cite{ritsch_cold_2013},
\begin{subequations}
\begin{equation}
\frac{d}{dt}\hat{a}_{\ell} = i \left[ \hat{H} / {\hbar}, \hat{a}_{\ell} \right] - \kappa \hat{a}_{\ell} + \hat{\Xi}_{\ell}(t),
\end{equation}
\begin{equation}
\frac{d}{dt}\hat{S}^{x, y, z}_{\ell} = i \left[ \hat{H}/ \hbar, \hat{S}^{x, y, z}_{\ell} \right].
\end{equation}
\end{subequations}
This approximation is typically employed in numerical simulations of atom-cavity systems, in which the system size is of orders $N \gtrsim 10^{4}$ \cite{ritsch_cold_2013}. Here, $\hat{\Xi}_{\ell}(t)$ is a Gaussian noise operator satisfying the relations $\langle \hat{\Xi}_{\ell}(t) \rangle = 0$ and $\langle \hat{\Xi}_{\ell}^{\dagger}(t)\hat{\Xi}_{\ell'}(t') \rangle = \kappa \delta(t - t')\delta_{\ell, \ell'}$. After evaluating the commutation relations between $\hat{H}$, $\hat{a}_{\ell}$ and $\hat{S}^{x, y, z}_{\ell}$, we approximate the system's dynamics by replacing the operators with complex variables, with $\hat{a}_{\ell}, \hat{\Xi}_{\ell}(t) \rightarrow a_{\ell}, \Xi_{\ell}(t) \in \mathbb{C}$ and $\hat{S}^{x, y, z}_{\ell} \rightarrow S^{x, y, z}_{\ell} \in \mathbb{R}$. Doing this results in a set of stochastic differential equations of the form,
\begin{subequations}
\label{eq:dlm_twa_eom}
\begin{equation}
\begin{split}
da_{\ell}^{\mathrm{R}} = & \left[ \omega a_{\ell}^{\mathrm{I}} - J \left( a_{\ell - 1}^{\mathrm{I}} + a_{\ell + 1}^{\mathrm{I}}\right) - \kappa a_{\ell}^{\mathrm{R}}    \right] dt \\
& + \sqrt{\frac{\kappa}{2}}dW_{1, \ell},
\end{split}
\end{equation}
\begin{equation}
\begin{split}
da_{\ell}^{\mathrm{I}} = &  - \left[  \omega a_{\ell}^{\mathrm{R}} + \frac{2\lambda}{\sqrt{N}}S^{x} - J \left(a_{\ell - 1}^{\mathrm{R}} + a_{\ell + 1}^{\mathrm{R}} \right) + \kappa a_{\ell}^{\mathrm{I}} \right]dt \\
& + \sqrt{\frac{\kappa}{2}} dW_{2, \ell},
\end{split}
\end{equation}
\begin{equation}
\frac{d}{dt}S^{x}_{\ell} = - \omega_{0}S_{\ell}^{y}, \quad \frac{d}{dt} S^{y}_{\ell} = \omega_{0}S^{x}_{\ell} - \frac{4\lambda}{\sqrt{N}}a_{\ell}^{\mathrm{R}}S^{z}_{\ell},
\end{equation}
\begin{equation}
 \frac{d}{dt}S^{z}_{\ell} = \frac{4\lambda}{\sqrt{N}}a_{\ell}^{\mathrm{R}}S^{y}_{\ell},
\end{equation}
\end{subequations}
where $a_{\ell} = a_{\ell}^{\mathrm{R}} + i a_{\ell}^{\mathrm{I}}$. Here, $W_{i}$ is a Wiener process satisfying the condition $\langle dW_{i} \rangle = 0$ and $\langle dW_{i}dW_{j} \rangle = \delta_{i, j} dt$ for $i \in \left\{1, 2 \right\}$.

\begin{figure}
\centering
\includegraphics[scale=0.355]{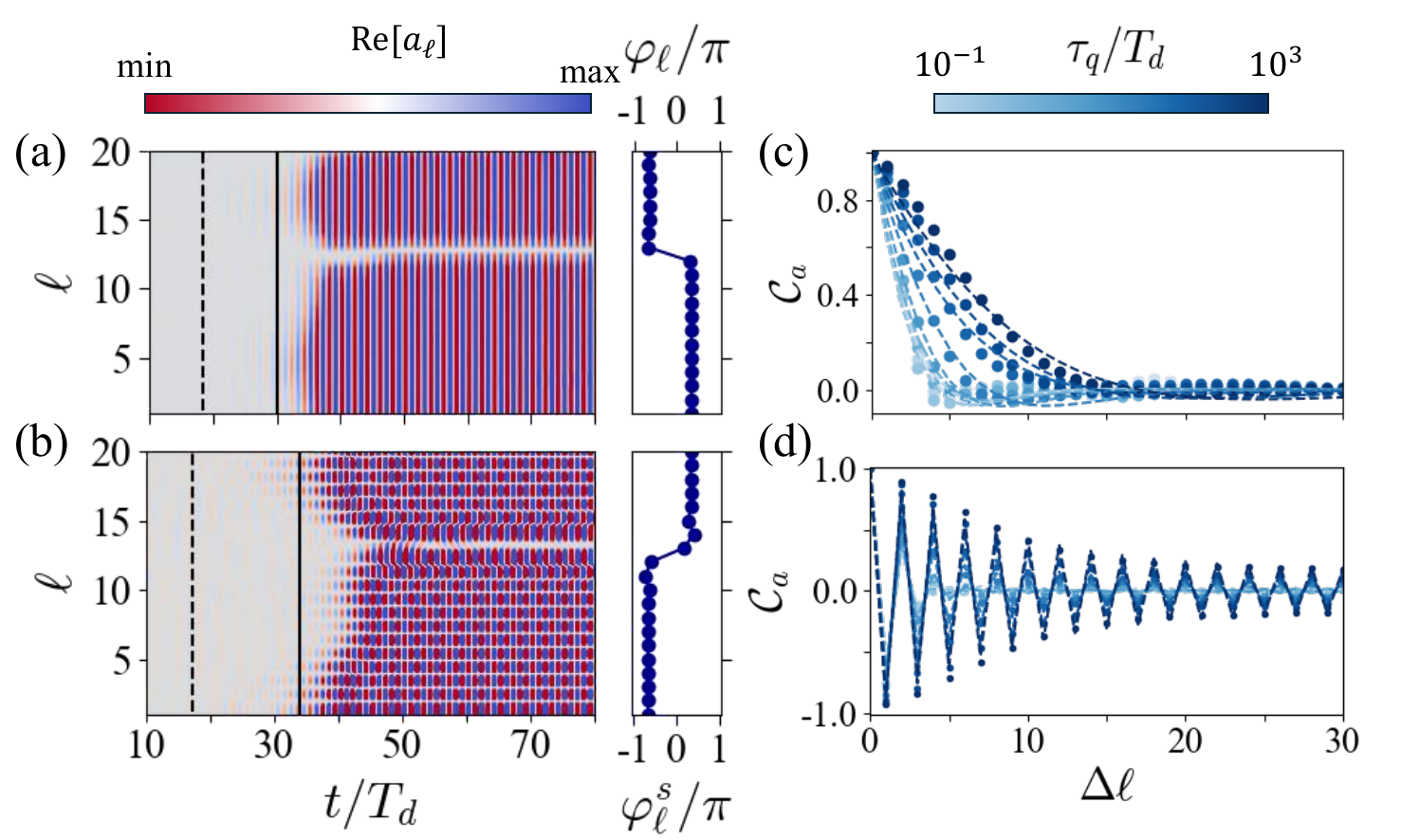}
\caption{(a)--(b) Left panels show the exemplary spatiotemporal dynamics of the DLM for (a) the FM-DTC and (b) the AFM-DTC and for $\tau_{q} = 50 T_{d}$. The solid lines correspond to the transition time $t_{p}$, while the dashed lines represent the critical time $t_{c}$. Right panels shows the steady state profile of $\varphi_{\ell}$ and $\varphi_{\ell}^{s}$ obtained at $t_{f}$. (c)--(d) Spatial correlation of the DLM at $t_{p}$ for (c) the FM-DTC and (d) the AFM-DTC, for different values of $\tau_{q}$. The dashed lines represent the best fit curve associated with the numerically obtained values of $C_{a}$. The parameters considered are $\left\{ \lambda_{0}, \kappa, J, N, M  \right\} = \left\{0.7 \lambda_{c}, 0.1\omega, 0.1\omega, 10^{4}, 100 \right\}$, with $\omega_{d} = 0.9\Omega$ for the FM-DTC and $\omega_{d} = 1.5\Omega$ for the AFM-DTC.}
\label{fig:dlm_dynamics}
\end{figure}

The crux of the truncated Wigner approximation lies in the sampling of multiple trajectories with different initial values and noise realizations to obtain the expectation value of the quantity we are interested in \cite{polkovnikov_phase_2010}. For the DLM, we initialize the bosonic modes in a vacuum state. This state can be numerically represented as a complex Gaussian number $a_{\ell}(t_{i}) = \left(\eta_{\ell}^{\mathrm{R}} + i \eta_{\ell}^{\mathrm{I}} \right) / 2$, where $\eta_{\ell}^{\mathrm{R, I}}$ satisfies the condition $\langle \eta_{\ell}^{i} \rangle = 0$ and $\langle \eta_{\ell}^{i}\eta_{\ell'}^{j} \rangle = \delta_{\ell, \ell'}\delta_{i, j} $ for $i, j \in \left\{ \mathrm{R}, \mathrm{I} \right\}$ \cite{olsen_numerical_2009}. Meanwhile, we consider a perturbed initial state of 
\begin{equation}
S^{x}_{\ell} = \epsilon \frac{N}{2}, \quad S^{y}_{\ell} = 0, \quad S_{\ell}^{z} = - \frac{N}{2}\sqrt{1 - \epsilon^{2}}, 
\end{equation}
with $\epsilon = 10^{-6}$, for the collective spins at each sites. Solving Eq.~\eqref{eq:dlm_twa_eom} for different noise realizations allows us to obtain a distribution for the transition delay $\hat{t}$ and defect number $n_{d}$, which we can then use to calculate the $\langle \hat{t} \rangle$ and $\langle n_{d} \rangle$ shown in the main text.

We present in Figs.~\ref{fig:dlm_dynamics}(a) and \ref{fig:dlm_dynamics}(b) single-trajectory dynamics of the DLM corresponding to the formation of FM-DTC and AFM-DTC, respectively, for $\tau_{q} = 50 T_{d}$. Similar to the SGM, the defects in the DLM manifest as phase slips in the stead state spatial profile of $\varphi_{\ell}$ ($\varphi_{\ell}^{s}$) for the FM-DTC (AFM-DTC). Likewise, as shown in Figs.~\ref{fig:dlm_dynamics}(c) and \ref{fig:dlm_dynamics}(d), the spatial correlation of the DLM also exhibits the same behavior as the spatial correlation of the SGM. In particular, for the FM-DTC, the spatial correlation exponentially decays to zero as $\Delta \ell$ increases, with the decay rate being dependent on the quench time. For the AFM-DTC, meanwhile, the spatial correlation exhibits oscillations that exponentially decay to a finite value.

\section{Extraction of critical point by scaling analysis}\label{sec:critical_point}

\begin{figure}
\centering
\includegraphics[scale=0.43]{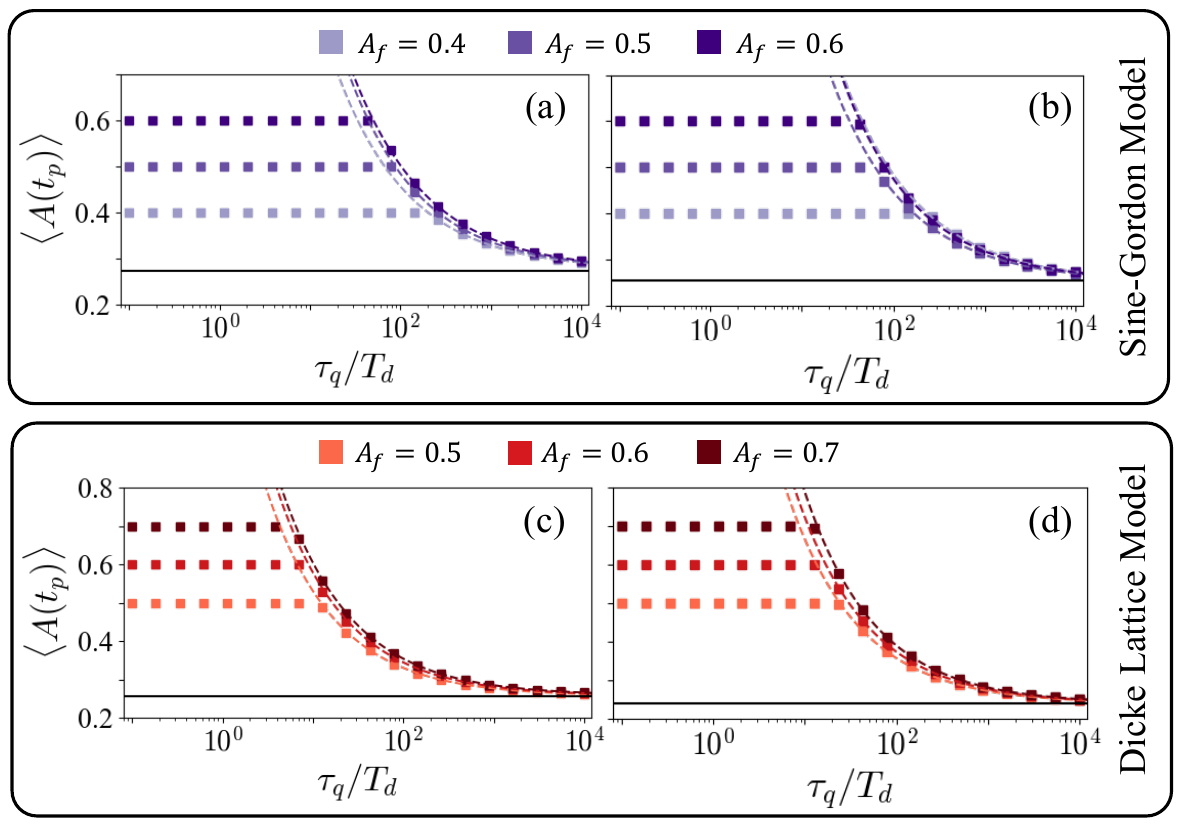}
\caption{(a)--(b) Ensemble average of the driving amplitude at the transition time $t_{p}$ of the SGM for (a) the FM-DTC and (b) the AFM-DTC. The same parameters used in Fig.~\ref{fig:kzm_scaling}(a) of the main text were used to construct (a) and (b). (c)--(d) Average $A(t_{p})$ for (c) the FM-DTC and (d) the AFM-DTC of the DLM, with the same parameters used in Fig.~\ref{fig:kzm_scaling}(b) of the main text. For all figures, the solid lines correspond to the extrapolated critical points, while the dashed lines represent the best-fit polynomial curves.}
\label{fig:amplitude_scaling}
\end{figure}

We identify $t_{c}$ by first obtaining the driving amplitude at the transition time, $A(t_{p})$, for every single trajectory. We then construct the scaling of the ensemble average of $A(t_{p})$, $\langle A(t_{p}) \rangle$, as a function of $\tau_{q}$. We show in Figs.~\ref{fig:amplitude_scaling}(a) and \ref{fig:amplitude_scaling}(b) the scaling of $\langle A(t_{p}) \rangle$ for the FM-DTC and AFM-DTC of the SGM, respectively, while we present in Figs.~\ref{fig:amplitude_scaling}(c) and \ref{fig:amplitude_scaling}(d) the same scaling for the FM-DTC and AFM-DTC of the DLM. Note that the range of $\tau_{q}$ in which $\langle A(t_{p}) \rangle$ remains constant corresponds to the breakdown regime of the KZM. Meanwhile, the range of $\tau_{q}$ in which $\langle A(t_{p}) \rangle$ follows a power-law behavior corresponds to the regime where the KZM is valid. We obtain the critical point by fitting the function $A(\tau_{q}) = a\tau_{q}^{-b} + c$ on the scaling of $\langle A(t_{p}) \rangle$. This is justified by how, in the KZM regime, it is predicted that 
\begin{equation}
\varepsilon(t_{p}) = \left[ A(t_{p}) - A_{c} \right] / A_{c} = (A_{0} / A_{c}) \tau_{q}^{-1/(1 + vz)},
\end{equation}
where $A_{0}$ is a proportionality constant \cite{del_campo_universality_2014}. Rewriting the scaling of $\varepsilon(t_{p})$ as 
\begin{equation}
A(t_{p}) = A_{0}\tau_{q}^{-1/(1 + vz)} + A_{c}
\end{equation}
makes it apparent that the fitting parameters $a$ and $b$ corresponds to the scaling coefficients and exponents of $\varepsilon(t_{p})$, respectively, while the fitting parameter $c$ corresponds to the critical point. We show the extrapolated critical points for the SGM and the DLM as solid horizontal lines in Fig.~\ref{fig:amplitude_scaling}. Finally, we obtain $t_{c}$ using Eq.~\ref{eq:critical_time}.

\section{Spatio-temporal defects in the strong noise limit}\label{sec:spacetime_defect}

\begin{figure}
\centering
\includegraphics[scale=0.42]{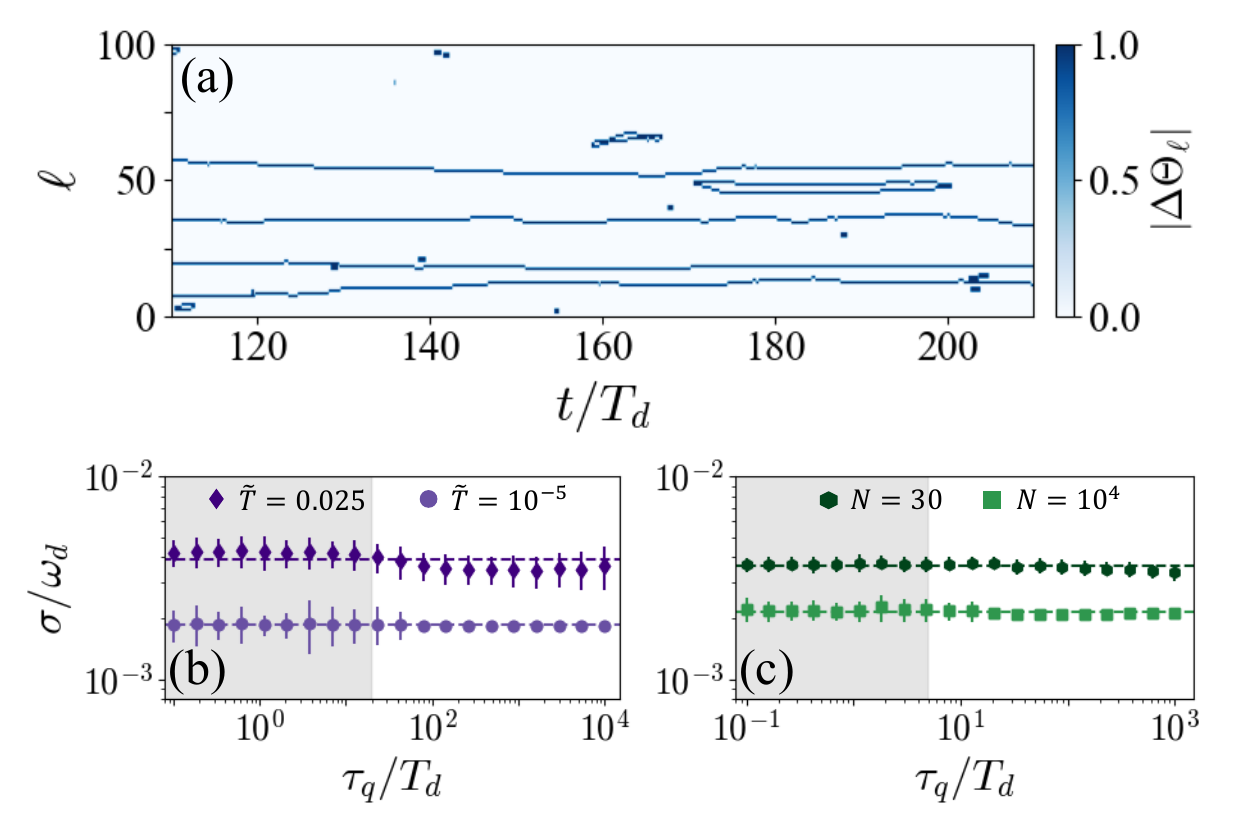}
\caption{(a) Exemplary defect dynamics of the DLM in the FM-DTC configuration, with $\tau_{q} = 10T_{d}$. (b)--(c) Temporal coherence of the FM-DTC for the (b) SGM and (c) DLM as a function of $\tau_{q}$ and noise strength. The shaded regions indicate the values of $\tau_{q}$ where the KZM breaks down, while the dashed lines represent the mean value of the data points relative to $\tau_{q}$. The parameters considered for the SGM are $\left\{ \gamma, g, \omega_{d}, A_{f}, M \right\} = \left\{ 0.1\Omega, 0.1\Omega^{2}, 1.9 \Omega, 0.6, 100 \right\}$, while we consider the parameters $\left\{ \lambda_{0}, \kappa, J, \omega_{d}, A_{f}, M \right\} = \left\{ 0.7\lambda_{c}, 0.1\omega, 0.1\omega,  0.9\omega, 0.7, 100 \right\}$ for the DLM. }
\label{fig:spatiotemporal_defect}
\end{figure}

In the weak noise limit considered in the main text, the fluctuations are weak enough to perturb the spatial defects appearing in the spatiotemporal dynamics of the DTC. This motivates the question of whether novel forms of defects can appear when the fluctuation becomes strong, and whether these defects follow a similar power-law scaling predicted by the KZM. We present in Fig.~\ref{fig:spatiotemporal_defect}(a) an exemplary spatiotemporal dynamics of the defects emerging in the FM-DTC of the DLM long after the quench has ended, for $N = 30$. We track the dynamics of the defects in the FM-DTC by obtaining the quantity
\begin{equation}
\Delta \Theta_{\ell} = \frac{1}{2}\left[  \mathrm{sgn}(\varphi_{\ell + 1}) - \mathrm{sgn}(\varphi_{\ell}) \right],  
\end{equation}
where $\mathrm{sgn}(\cdots)$ is the signum function. Notably, aside from the spatial defects appearing as line defects in the spatiotemporal dynamics of the DLM, we also observe spatiotemporal defects appearing as closed loops in Fig.~\ref{fig:spatiotemporal_defect}(a). These defects correspond to small regions in the lattice jumping towards the other degenerate state of the DTC due to strong fluctuations. These regions, however, are short-lived, as evinced by the small area occupied by these defects in Fig.~\ref{fig:spatiotemporal_defect}(a).

To identify whether these spatiotemporal defects follow the KZM, we quantify the temporal coherence of the DTC oscillation at each site by calculating the site-averaged power spectrum of a generic order parameter $O_{\ell, \mathrm{R}}(t)$, 
\begin{equation}
\mathbb{O}(\omega_{r}) = \frac{ \sum_{k} \left< |O_{k, \mathrm{R}} (\omega_{r})|^{2} \right>}{\sum_{k} \left< |O_{k, \mathrm{R}}(\omega_{r} = 0)|^{2} \right> },
\end{equation}
where $O_{k, \mathrm{R}}(\omega_{r})$ is the space-time Fourier transform of $O_{\ell, R}$ defined as
\begin{equation}
\label{eq:spacetime_fft}
O_{k, \mathrm{R}}(\omega_{r}) = \frac{1}{\sqrt{2\pi M}} \sum_{\ell = 1}^{M} \int_{t_{s, i}}^{t_{s, f}} dt \left [ e^{i \left( k\ell + \omega_{r}t \right)} O_{\ell , \mathrm{R}}(t) \right].
\end{equation}
The integral in Eq.~\eqref{eq:spacetime_fft} is evaluated from some initial scan time $t_{s, i}$ to some final scan time $t_{s, f}$. Assuming that $\mathbb{O}(\omega_{r})$ can be approximated as a Gaussian function, $\mathbb{O}(\omega_{r}) = \exp\left( - \omega^{2}_{r} / 2\sigma^{2} \right)$, the temporal coherence of the DTC is then characterized by inverse of $\sigma$. A smaller $\sigma$ indicates longer temporal coherence for the DTC, while a larger value indicates the presence of spatiotemporal defects. In this interpretation, $\sigma^{-1}$ is thus the temporal analog of the correlation length $\xi$ calculated in the main text.

We now present in Figs.~\ref{fig:spatiotemporal_defect}(b) and \ref{fig:spatiotemporal_defect}(c) the scaling of $\sigma$ as a function of $\tau_{q}$ for different values of $\tilde{T}$ and $N$, respectively. We can observe that for all $\tau_{q}$ considered, $\sigma$ remains constant both in the weak and strong noise limit. These results demonstrate that the spatiotemporal defects observed in the strong noise limit are independent of the correlation build-up induced by ramping the system across the critical point. They rather originate from the fluctuations, and thus only become relevant in the strong noise limit. This behavior is consistent with the behavior observed in the SGM when driven across a finite-temperature phase transition \cite{yao_classical_2020}.

\bibliography{dtc_kzm_main_ref.bib}

\end{document}